\documentclass[superscriptaddress,showpacs,amssymb,10pt,aps,prd,reprint,longbibliography,footinbib]{revtex4-1}

\usepackage{graphicx}
\usepackage{bm}
\usepackage{rotating}
\usepackage{array}

\usepackage{graphicx,epsfig,amssymb,times}
\usepackage{amsmath,amsfonts}
\usepackage{bm}
\usepackage{epstopdf}
\usepackage[linktocpage,colorlinks]{hyperref}
\usepackage[caption=false]{subfig}
\usepackage[usenames]{color}
\usepackage{float}
\usepackage{natbib}
\usepackage{ulem}
\usepackage{cancel}
\usepackage{verbatim}
\usepackage{enumitem}
\usepackage[utf8]{inputenc}
\definecolor{coolblack}{rgb}{0.0, 0.18, 0.39}
\definecolor{darkred}{rgb}{0.5,0,0}
\definecolor{darkgreen}{rgb}{0,0.5,0}
\definecolor{darkblue}{rgb}{0,0,0.5}
\definecolor{lapislazuli}{rgb}{0.15, 0.38, 0.61}
\definecolor{venetianred}{rgb}{0.78, 0.03, 0.08}
\definecolor{bleudefrance}{rgb}{0.19, 0.55, 0.91}
\definecolor{dogwoodrose}{rgb}{0.84, 0.09, 0.41}
\hypersetup{colorlinks=true, citecolor=darkblue, linkcolor=darkblue,
urlcolor = darkblue}

\def\be{\begin{equation}}
\def\ee{\end{equation}}

\def\({\left(}
\def\){\right)}
\def\[{\left[}
\def\]{\right]}

\def\e{\begin{equation}}
\def\q{\end{equation}}
\def\m{\begin{eqnarray}}
\def\n{\end{eqnarray}}

\newcommand{\bea}{\begin{eqnarray}}
\newcommand{\eea}{\end{eqnarray}}
\newcommand{\ben}{\begin{enumerate}}
\newcommand{\een}{\end{enumerate}}
\newcommand{\bi}{\begin{itemize}}
\newcommand{\ei}{\end{itemize}}

\def\ga{\mathrel{\raise.3ex\hbox{$>$\kern-.75em\lower1ex\hbox{$\sim$}}}}
\def\la{\mathrel{\raise.3ex\hbox{$<$\kern-.75em\lower1ex\hbox{$\sim$}}}}

\def\be{\begin{equation}}
\def\ee{\end{equation}}

\def\I_M{{I_{\scriptscriptstyle M\times M}}}

\def\be{\begin{equation}}
\def\ee{\end{equation}}
\def\bea{\begin{eqnarray}}
\def\eea{\end{eqnarray}}

\begin{document}

\title{\large Constraints on Multicomponent Dark Energy from Cosmological Observations}

\author{Ke Wang}\email{wangkey@lzu.edu.cn}
\affiliation{Institute of Theoretical Physics \& Research Center of Gravitation,\\ Lanzhou University, Lanzhou 730000, China}

\author{Lu Chen}\email{Corresponding author: chenlu@sdnu.edu.cn}
\affiliation{School of Physics and Electronics, \\Shandong Normal University, Jinan 250014, China\\}
 
 \date{\today}

\begin{abstract}
Dark energy (DE) plays an important role in the expansion history of our universe.
But we only got limited knowledge about its nature and properties after decades of study.
In most numerical researches, DE is usually considered as a dynamical whole.
Actually, multicomponent DE models can also explain the accelerating expansion of our universe,
which is accepted theoretically but lack of numerical researches.
We try to study the multicomponent DE models from observation by constructing $w_n$CDM models.
The total energy density of DE is separated into $n$ ($n=2,3,5$) parts equally and every part has a constant EOS $w_i$ ($i=1,2...n$).
We modify the Friedmann equation and the parameterized post-Friedmann description of DE, then put constraints on $w_i$s from Planck 2018 TT,TE,EE$+$lowE$+$lensing, BAO data and PANTHEON samples. 
The multicomponent DE models are favoured if any $w_n$CDM model is preferred by observational data and there is no overlap between the highest and lowest values of $w_i$s. 
We find the data combination supports the $w_n$CDM model when $n$ is small and the $w_2$CDM model is slightly preferred by $\Delta \chi^2_{\text{min}} = \Delta \text{AIC} =\Delta \text{BIC} = -2.48$ over the CPL model, but the largest value of $w_i$ overlaps the smallest one. With larger $n$, the maximum and minimum of $w_i$s do not overlap with each other, but $\chi^2_{\text{min}}$ and AIC also increase.
In brief, we find no obvious evidence that DE is composed of different components.
\end{abstract}

\maketitle
\section{Introduction} 
\label{sec:intro}
The concept of dark energy (DE) has been widely accepted by many physicists since the discovery of cosmic accelerating expansion in 1998~\cite{Riess:1998cb,Perlmutter:1998np}.
Lots of methods are proposed to study the properties of DE.
The distance measurements are used to detect DE, such as the baryonic acoustic oscillation (BAO) observation~\cite{Cole:2005sx,Ross:2014qpa,Beutler:2011hx,Alam:2016hwk,Ata:2017dya,Carter:2018vce,Abbott:2017wcz,Agathe:2019vsu,Cao:2020jgu} and surveys on Type Ia supernovae (SNIa)~\cite{Conley:2011ku,Suzuki:2011hu,Scolnic:2017caz}.
The formation of large scale structure~\cite{Abate:2012za,Mandelbaum:2018ouv,Korytov:2019xus,Lochner:2018boe} is also influenced by DE significantly due to its negative pressure.
DE, as the largest proportion of the total energy density in today's universe, also leaves footprints on the cosmic microwave background (CMB)~\cite{Komatsu:2008hk,Ade:2013zuv,Ade:2015xua,Aghanim:2018eyx}.
However, the nature of DE is still a puzzle through decades of researches.

Theoretically, DE is considered as the cosmological constant firstly~\cite{Peebles:2002gy}. After that, fluids or fields, especially scalar fields~\cite{Peebles:1987ek,Ratra:1987rm,Novosyadlyj:2010pg,Johnson:2020gzn,MohseniSadjadi:2020jmc,Kase:2020hst}, also be considered as candidates of DE.
Numerically, DE is usually considered as to be a cosmological constant or dynamical models with its equation of state (EOS) $w$ varies with redshift $z$.
For the simple cases, $w$ is constant -1 in the base $\Lambda$CDM model and it is a free parameter in the $w$CDM model. 
In the CPL model (also named the $w_0 w_a$CDM model)~\cite{Linder:2002et,Chevallier:2000qy} and other dynamical DE models, $w$ is assumed to be functions of $z$. They have been well investigated in the previous works~\cite{Ade:2015xua,Aghanim:2018eyx,Vagnozzi:2018jhn,Chen:2018dbv,Chen:2017ayg,Huang:2015vpa}, both theoretically and numerically.
In fact, there is another possibility that DE is composed of different components, such as multi-field models~\cite{Aghanim:2018eyx,DeFelice:2010aj,Vardanyan:2015oha,Akrami:2020zfz}.
Previous works indicate we can not distinguish the multi-field DE from single-field DE unless perturbations are considered.
DE perturbations can cluster and leave footprints on structure formation under some circumstances, which is the way we judge whether DE is a whole or not.
However, there are few numerical works aiming to study the constitution of DE starting from observational data. 
In our work, we try to study models with multicomponent DE from cosmological observation.
Assuming DE only has gravitational interaction or it is minimally coupled to other components of universe, we separate today's total DE energy density into $n$ parts equally (hereafter $w_n$CDM model) and reconsider the cosmic expansion history, including the background and perturbation evolutions. 
{\it Notice that ``$n$ parts" dose not mean DE is composed of $n$ kinds of candidates.} Actually, it is just a numerical separation artificially.
Then we denote the constant EOSs of different parts as $w_i$s ($i=1,2\dots n$) and put constraints on them from observational data combination. 
We can not tell the $i$th and the $j$th ($j=1,2\dots n$, $j\neq i$) parts belong to different components of DE if $w_i$ and $w_j$ overlap with each other conservatively.
However, if any $w_n$CDM model satisfies both of the following conditions: (i) there is no overlap between the EOSs of any two parts at some confidence level, (ii) the model is favoured by observational data, we will reach a conclusion that DE is composed of more than one candidate.

This paper is organized as follows.
In section \ref{sec:ppf}, we sketch out the background and perturbation evolutions of the $w_n$CDM models.
We utilize the CMB data~\cite{Aghanim:2018eyx}, BAO data~\cite{Alam:2016hwk,Carter:2018vce,Abbott:2017wcz,Agathe:2019vsu,Ata:2017dya,Cao:2020jgu} and SNIa measurement~\cite{Scolnic:2017caz} to constrain different DE models and show our results in section \ref{sec:results}.
Finally, a brief summary are included in section \ref{sec:sum}.

\section{parameterized post-Friedmann description of the $w_n\rm{CDM}$ models}
\label{sec:ppf}

In the $w_n$CDM models, we divide today's DE energy density into $n$ parts equally.
Then the total energy density of our universe is 
\m
\label{eq:rho}
\rho_{tot}(a)= \rho_T(a)+ \rho_{de} (a)=\rho_T(a)+ \sum_{i=1}^{n} \rho_{de,i}(1) a^{-3-3w_i}.
\n
Here $\rho_{de,i}(1)$ and $w_i$ are today's energy density and the constant EOS of the $i$th DE part respectively. $\rho_{de} $ is the total DE energy density. The subscript $T$ denotes all the other components excluding DE. $a=1/(1+z)$ is the scale factor.
The Friedmann equation is modified with Eq.~(\ref{eq:rho}).
Actually, considering only the background evolution, we can never distinguish a single-field DE from multifield DE because there always exists a single field leading to any observed Hubble parameter $H(z)$~\cite{Vardanyan:2015oha}. 
But we still modify this part for safe and integrity.

Then the evolution of DE perturbation should be dealt with.
To cross the phantom divide line, the evolution of DE perturbation is described with the parameterized post-Friedmann (PPF) description~\cite{Hu:2008zd,Fang:2008kc,Fang:2008sn,Grande:2008re,Wang:2012uf}.
Note that we take phantom divide into account though current data don't support phantom divide crossing if they are used together~\cite{Park:2018bwy}.
There are three reasons for doing so: i).the previous conclusion is acquired under the condition that DE is treated as a whole; ii).some models of multicomponent DE with at least one non-canonical phantom component can explain phantom divide crossing at recent redshifts~\cite{Nesseris:2006er}; iii).there are uncertainties in numerical analysis with different data combinations.
In the synchronous gauge, the perturbations of energy density and momentum of DE satisfy the following modified equations,
  \m
  \rho_{de} \delta_{de} = -3 \rho_{de}^{w} \dfrac{v_{de}}{k_H} -\dfrac{c_K k_H^2 H^2}{4 \pi G} \Gamma,
  \n
  \m
  \rho_{de}^{w} v_{de}   = \rho_{de}^{w} v_T - \dfrac{k_H^2 H^2}{4 \pi G F} \[ S-\Gamma-\dfrac{\dot{\Gamma}}{H}  \].
  \n
Here $\rho_{de}^{w} $ is defined as 
  \m
  \rho_{de}^{w} \equiv \sum_{i=1}^{n} \rho_{de,i}(1)a^{-3-3w_i} (1+w_i).
  \n
$\delta_{de}= \delta \rho_{de}/\rho_{de}=\delta_{de,i}= \delta \rho_{de,i}/\rho_{de,i}$ is the density perturbation of total DE and $\delta \rho_{de}=\sum_{i=1}^{n}\delta \rho_{de,i}$.
Note that it is reasonable to add them up directly if we assume different parts of DE are minimally coupled with each other in our models.
$v$ denotes velocity and $v_{de}=v_{de,i}$. $G$ is Newton's constant. $H$ is Hubble parameter.
$c_K$ is related to the background curvature of our universe. For a spatial flat universe, we have $c_K=1$.
$k_H = k/aH$, where $k$ is the wave number in Fourier space.
Overdot means the differentiation over cosmic time.
Besides,  
  \m
  F=1+\dfrac{12 \pi G a^2}{k^2 c_K} (\rho_T +p_T),
  \n
 \m
 S=\dfrac{4 \pi G}{H^2}  \dfrac{(v_T+k \alpha)}{k_H} \rho_{de}^{w},
 \n  
where $\alpha = a(\dot{h}+6\dot{\eta})/2k^2$, $h$ and $\eta$ is the metric perturbations in the synchronous gauge.
PPF description provides a well approximation for minimally coupled scalar field DE models and many smooth DE models.
Therefore, DE ought to be relatively smoother than matter inside a transition scale $c_s k_H =1$,
\m
\rho_{de} \delta_{de} \ll \rho_{T} \delta_{T}.
\n
In order to satisfy this condition, we have the following differential equation for $\Gamma$
\m
(1+c_{\Gamma}^2 k_H^2) \[ \Gamma +c_{\Gamma}^2 k_H^2 \Gamma +\dfrac{\dot{\Gamma}}{H} \] =S.
\n
For any given evolution of $\Gamma$, there is a specific evolution of perturbations of DE. And $c_{\Gamma}=0.4 c_s$ for the evolution of scalar field models.

\section{Results}
\label{sec:results}
We refer to CAMB$+$CosmoMC packages~\cite{Lewis:1999bs,Lewis:2002ah,Lewis:2013hha}   and  use the data combination of CMB, BAO and SNIa measurements to constrain the EOSs of different DE models.
Concretely, we use Planck2018 TT,TE.EE$+$lowE$+$lensing~\cite{Aghanim:2018eyx}, the BAO measurements at $z=0.122,0.38, 0.51, 0.61, 0.81, 1.52,2.34$~\cite{Alam:2016hwk,Carter:2018vce,Abbott:2017wcz,Agathe:2019vsu,Ata:2017dya} (which are summerized in Ref.~\cite{Cao:2020jgu}), as well as the PANTHEON samples~\cite{Scolnic:2017caz}.

Based on the previous discussion in Sec.~\ref{sec:ppf}, we divide today's total density of DE into $n=$2, 3 or 5 parts averagely and modify both the background and perturbation evolutions of DE in each model.
It is noted here we also modify the ``halofit" code~\cite{Mead:2016zqy} included in CAMB package, which models the non-linear matter power spectrum resulting from parameterized DE models.
We run CosmoMC with ``action=0" to get the marginalized errors and ``action=2" to get the minimum values of $\chi^2$.
In $w_2$CDM model, there are eight free parameters needed to be fitted: $\{\Omega_b h^2,  \Omega_c h^2, 100\theta_{\text{MC}}, \tau, \ln(10^{10}) A_s, n_s,$ $w_1, dw_2\}$.
In the case of $w_{3}$CDM model, $dw_3$ is added and there are nine free parameters.
Another two free parameters $\{ dw_4, dw_5\}$ are added in the $w_{5}$CDM model.
Six of them are parameters in the base $\Lambda$CDM model. $\Omega_b h^2$ and $\Omega_c h^2$ are today's density of baryonic matter and cold dark matter respectively,
$100\theta_{\text{MC}}$ is 100 times the ratio of the angular diameter distance to the large scale structure sound horizon,
$\tau$ is the optical depth, $n_s$ is the scalar spectrum index, and $A_s$ is the amplitude of the power spectrum of primordial curvature perturbations.
$w_1$ is the largest one among all the values of $w_i$ ($i=1,2...n$).
$dw_j (j=2,3,4,5)$ is the difference between the EOSs of two adjacent values.
In other words, EOSs of different DE parts are listed from top to bottom as below: $w_1, w_2=w_1-dw_2, w_3=w_1-dw_2-dw_3$ and so forth.
The ranges of $w_1$ is set to be $\[-10,10\]$ and $dw_j \in \[0,3 \]$.
For comparison, the $\Lambda$CDM, $w$CDM and $w_0w_a$CDM models are also carried out.

\linespread{1.5}
\begin{table*}
\label{tab:de}
\caption{Constraints on the cosmological parameters in different DE models from the combination data of Planck 2018 TT,TE,EE$+$lowE$+$lensing, BAO data and PANTHEON samples. The first set of error bars indicates the 68$\%$ limits and the second set in parentheses reflects the 95$\%$ limits. Note that we get their constraints with ``action=0" in CosmoMC packages, but we set ``action=2" to get more accurate values of $\chi^2_{\text{min}}$ and $\text{AIC}$.}
\begin{tabular}{p{2.7cm} <{\centering}p{4.0cm}<{\centering} p{4.0cm}<{\centering}  p{4.cm}<{\centering} }
\hline
\hline
                 & $\Lambda$CDM model & $w$CDM model  & $w_{0}w_a$CDM model \\
\hline
$\Omega_b h^{2}$   & $0.02243\pm 0.00014 (\pm 0.00027)$ & $0.02240\pm 0.00014 (_{-0.00027}^{+0.00028})$ & $0.02238\pm 0.00014 (\pm 0.00027)$   \\

$\Omega_c h^{2}$        & $0.1192\pm 0.0009 (\pm 0.0017)$   & $0.1196\pm 0.0010 (_{-0.0021}^{+0.0020})$  & $0.1199\pm 0.0011 (\pm 0.0021)$  \\

$100\theta_{MC}$& $1.04101\pm 0.00029 (^{+0.00056}_{-0.00057})$& $1.04097\pm 0.00029(\pm 0.00059) $& $1.04092\pm 0.00030 (\pm 0.00059) $  \\

$\tau$    & $0.057 _{-0.008}^{+0.007} (_{-0.014}^{+0.015})$  & $0.055^{+0.007}_{-0.008} (_{-0.014}^{+0.015})$& $0.053\pm 0.007 (\pm 0.015)$      \\
$\ln(10^{10}A_s)$     &$3.047^{+0.014}_{-0.015} (_{-0.027}^{+0.030})$& $3.045_{-0.015}^{+0.014}(_{-0.028}^{+0.029})$     &$3.042\pm 0.014(_{-0.029}^{+0.030})$    \\
$n_s$ & $0.9669_{-0.0035}^{+0.0036}(_{-0.0068}^{+0.0069}) $     & $0.9661\pm 0.0038(^{+0.0076}_{-0.0075}) $& $0.9653\pm 0.0039 (_{-0.0079}^{+0.0078}) $    \\
\hline
$H_0$ [km s$^{-1}$ Mpc$^{-1}$]  &$67.71\pm 0.40(_{-0.77}^{+0.78}) $ &$68.29\pm 0.81(_{-1.59}^{+1.60}) $ &$68.23_{-0.82}^{+0.81}(_{-1.61}^{+1.62}) $ \\
\hline
 $w(w_0)$  & - &$-1.0247_{-0.0316}^{+0.0313}(\pm 0.0627)  $      & $-0.9507_{-0.0774}^{+0.0773}  (_{-0.1531}^{+0.1569})  $         \\
 $w_a$&-&-  &  $-0.3011_{-0.2743}^{+0.3165} (_{-0.5971}^{+0.5781})  $      \\
\hline
$\chi^2_{\text{min}}$      &  3817.640             & 3817.106      &  3819.502 \\
 $\text{AIC}$  &  3829.640           & 3831.106  &  3835.502   \\

\hline
\hline
                 & $w_{2}$CDM model & $w_{3}$CDM model  & $w_{5}$CDM model \\
\hline
$\Omega_b h^{2}$   & $0.02240\pm 0.00014 (\pm 0.00027)$ & $0.02241\pm 0.00014 (\pm 0.00027)$ & $0.02242\pm 0.00014 (\pm 0.00027)$   \\

$\Omega_c h^{2}$        & $0.1195\pm 0.0010 (\pm0.0020)$   & $0.1195\pm 0.0010 (_{-0.0021}^{+0.0020})$  & $0.1194\pm 0.0010 (\pm0.0020)$  \\

$100\theta_{MC}$& $1.04098\pm 0.00030 (^{+0.00058}_{-0.00059})$& $1.04098\pm 0.00030(\pm 0.00060) $& $1.04100_{-0.00029}^{+0.00030} (\pm 0.00058) $  \\

$\tau$              & $0.056 _{-0.008}^{+0.007} (_{-0.014}^{+0.015})$     & $0.056^{+0.007}_{-0.008} (_{-0.014}^{+0.015})$    & $0.057^{+0.007}_{-0.008} (_{-0.014}^{+0.016})$      \\
$\ln(10^{10}A_s)$     &$3.046 \pm 0.014 (_{-0.027}^{+0.029})$& $3.046\pm 0.014 (_{-0.028}^{+0.029})$     &$3.048_{-0.015}^{+0.014} (_{-0.028}^{+0.030})$    \\
$n_s$                   & $0.9661\pm 0.0039 (_{-0.0076}^{+0.0077}) $     & $0.9662\pm 0.0039 (\pm 0.0076) $        & $0.9666\pm0.0039 (_{-0.0076}^{+0.0077}) $    \\
\hline
$H_0$ [km s$^{-1}$ Mpc$^{-1}$]     &$68.30\pm 0.80(_{-1.56}^{+1.60}) $ &$68.31_{-0.81}^{+0.82}(_{-1.60}^{+1.62}) $ &$68.49\pm0.82(_{-1.60}^{+1.65}) $ \\
\hline
 $w_1$  & $-0.9305_{-0.0787}^{+0.0612}  (_{-0.1229}^{+0.1327})$ &$-0.8600_{-0.0860}^{+0.0767}  (_{-0.1430}^{+0.1496})  $      & $-0.7900_{-0.0771}^{+0.0780}  (_{-0.1480}^{+0.1498})  $         \\
 $w_2$&$-1.1592_{-0.0647}^{+0.1386}(_{-0.2361}^{+0.1880})$&$-1.0223_{-0.0874}^{+0.0896}(_{-0.1677}^{+0.1629})  $        &  $-0.9192_{-0.0709}^{+0.0817} (_{-0.1512}^{+0.1480})  $                            \\
   $w_3$                       &   -            & $-1.2356_{-0.0832}^{+0.1670} (_{-0.2857}^{+0.2355}) $     &   $-1.0433_{-0.0760}^{+0.1020} (_{-0.1842}^{+0.1720}) $                  \\
    $w_4$  &  -  &- &  $ -1.2097_{-0.0934}^{+0.1563}  (_{-0.2714}^{+0.2389})  $  \\
  $w_5$   & - & -  &   $-1.5608_{-0.1446}^{+0.3731} (_{-0.6871}^{+0.5252})  $    \\
\hline
$\chi^2_{\text{min}}$      &  3817.022             & 3816.911      &  3817.139 \\
 $\text{AIC}$  &  3833.022            & 3834.911  &  3839.139   \\
\hline
\hline
\end{tabular}
\end{table*}

 \linespread{1.0}

\begin{figure}
\centering
\begin{minipage}[c]{1\linewidth}
\centering
\includegraphics[scale=0.62]{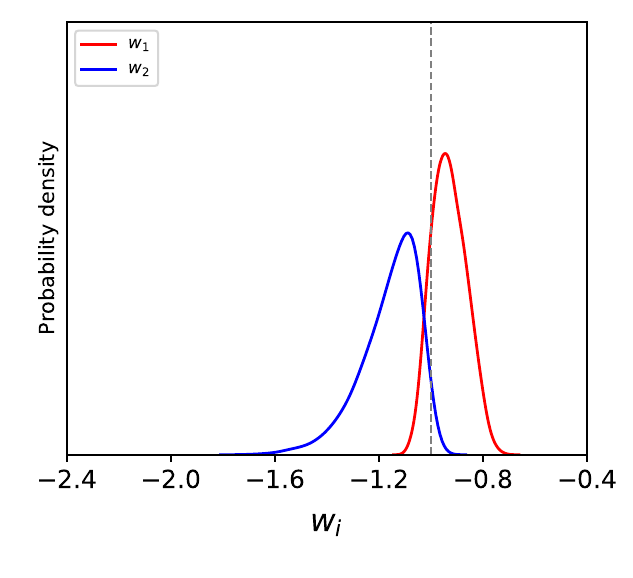}
\end{minipage}%

\begin{minipage}[c]{1\linewidth}
\includegraphics[scale=0.65]{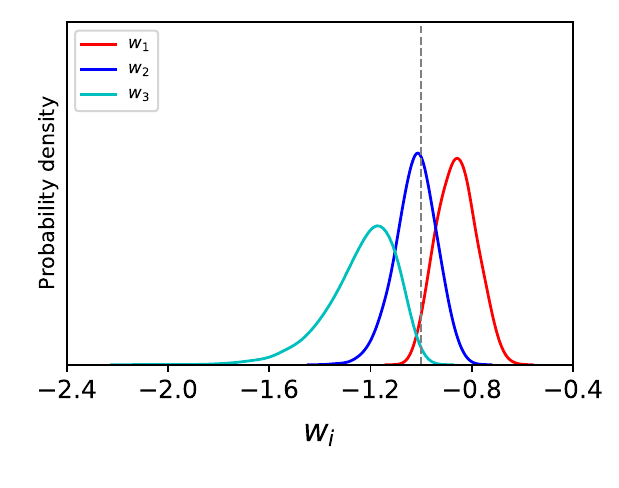}
\end{minipage}%

\begin{minipage}[c]{1\linewidth}
\includegraphics[scale=0.65]{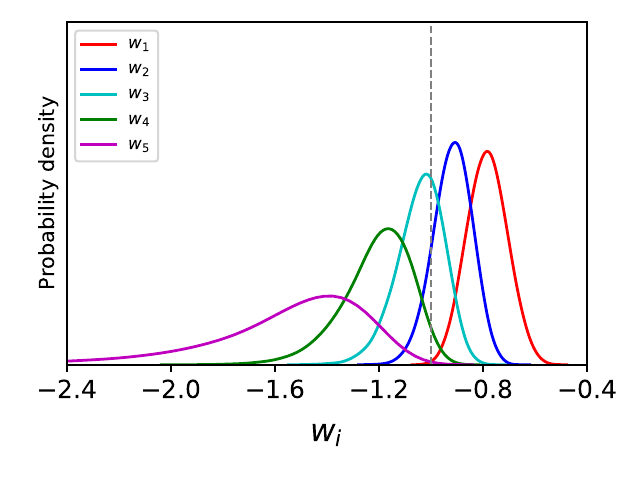}
\end{minipage}%
\caption{The probability densities of $w_i$ ($i=1,2...n$) in the $w_{n}$CDM ($n=2,3,5$) model. From top to bottom are $w_2$CDM, $w_3$CDM and $w_5$CDM model. The grey dashed lines denote the phantom divide $w_i=-1$.}
\label{fig:1d}
\end{figure}

The main results are shown in Tab.~I.
We show the 68$\%$ and 95$\%$ limits for the parameters in above six models.
Besides, the probability densities of $w_i$ in three $w_{n}$CDM ($n=2,3,5$) model are illustrated in Fig.~\ref{fig:1d} vividly.
Notice that constraints on $w_i$ are well around $-1$,
but much weaker limits are acquired when $w_i$ is much lower than -1 and there is a long left tail for $w_n$ in the $w_n$CDM model.
This is because DE component with $w_i \ll -1$ has little influence on the relatively early universe due to the suppression of scale factor in Eq.~(\ref{eq:rho}).
Our constraints and error bars on the six parameters of base $\Lambda$CDM model consist well with the values from Planck 2018 TT,TE,EE$+$lowE$+$lensing and BAO measurements given by Planck collaboration~\cite{Aghanim:2018eyx}.
In the $w_2$CDM model, the highest value of EOSs is $w_1=0.9305^{+0.0612}_{-0.0787}$ at 68$\%$ C.L. and $w_1=-0.9305_{-0.1229}^{+0.1327}$ at 95$\%$ C.L., while the lowest one is $w_2=-1.1592^{+0.13886}_{-0.0647}$ at 68$\%$ C.L. and $-1.1592_{-0.2361}^{+0.1880}$ at 95$\%$ C.L..
Although there is no overlap in the allowed ranges of 1$\sigma$, the overlap appears when their confidence level is around 2$\sigma$.
We reach a similar conclusion in the $w_{3}$CDM model.
But it is not the case for the $w_5$CDM model.
We get the highest EOS as $w_1=-0.7900_{-0.0771}^{+0.0780}$ at 68$\%$ C.L., $-0.7900_{-0.1480}^{+0.1498} $ at 95$\%$ C.L. and the lowest value as $w_5=-1.5608_{-0.1446}^{+0.3731} $ at 68$\%$ C.L., $-1.5608_{-0.6871}^{+0.5252} $ at 95$\%$ C.L..
It seems that the $w_5$CDM model prefers multicomponent DE because $w_1$ and $w_5$ show no overlap with each other at 2$\sigma$. 
However, values of $\chi^2_{\text{min}}$ and AIC are not conducive to $w_5$CDM model.
As Tab.~I shows, we can sort these models by $\chi^2_{\text{min}}$ from the smallest to the largest as $\{w_3\text{CDM}, w_2\text{CDM}, w\text{CDM}, w_5\text{CDM}, \Lambda \text{CDM}, w_0w_a\text{CDM}\}$.
Considering various numbers of free parameters in different models, we compare their Akaike information criterion (AIC) values with $\text{AIC}=\chi^2_{\text{min}}+2k$, where $k$ is the number of free parameters~\cite{aic1,aic2,Gong:2007se}.
So we put them in the order as $\{\Lambda \text{CDM},w\text{CDM},w_2\text{CDM}, w_3\text{CDM}, w_0w_a\text{CDM}, w_5\text{CDM}\}$.
However, the $w_3$CDM and $w_5$CDM models are extremely disfavored if we drop the AIC and use the Bayesian information criterion (BIC) given by $\text{BIC}=\chi^2_{\text{min}} +k \ln N$~\cite{Schwarz:1978tpv,Liddle:2004nh,Biesiada:2007um,Kurek:2007tb}, where $N$ denotes the number of data and has a very high value over 1000 in our work. 
The $w_2$CDM model is preferred over the $w_0w_a$CDM model by $\Delta \chi^2_{\text{min}} = \Delta \text{AIC} = \Delta \text{BIC} =-2.48$.
All in all, none of the $w_n$CDM ($n=2,3,5$) models satasfies the two conditions in Sec.~\ref{sec:intro}.
At this point, we find no obvious evidence of multicomponent DE.

Hereinafter, we focus on the $w$CDM, $w_0w_a$CDM and $w_2$CDM models. 

\begin{figure}[]
\begin{center}
\includegraphics[scale=0.29]{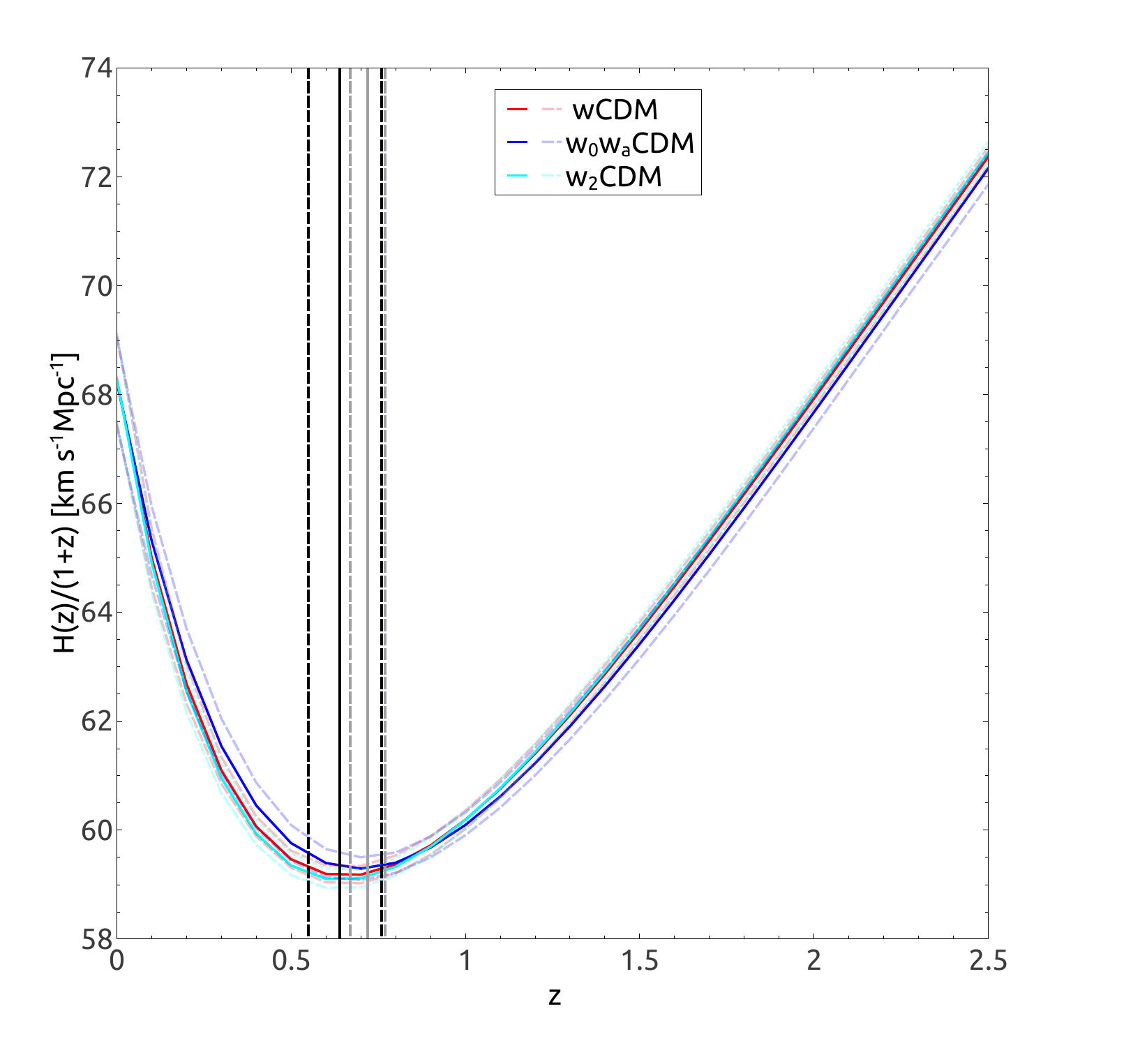}
\end{center}
\caption{ Comoving Hubble parameter $H(z)/(1+z)$ as a function of redshift $z$. The red, blue and cyan lines represent $H(z)/(1+z)$ in the $w$CDM, $w_0w_a$CDM and $w_2$CDM model respectively. Their mean values and  68$\%$ limits are denoted with solid and dashed lines respectively. Here the vertical black lines indicate the transition redshift $z_{\text{T}}=0.64^{+0.12}_{-0.09}$ from combined data of SNIa, BAO and Cosmic Chronometers (CC) data at low redshifts~\cite{Haridasu:2018gqm}. The vertical grey lines denote $z_{\text{T}}=0.72\pm 0.05$ from 38 measurements of $H(z)$ between redshifts $0.07 \leq z \leq 2.36$~\cite{Farooq:2016zwm}.
}
\label{fig:Hz}
\end{figure}

In Tab.~I, values of Hubble constant $H_0$ are listed. In the $w_2$CDM model, we have $H_0=68.30\pm 0.80$ km s$^{-1}$ Mpc$^{-1}$ at 68$\%$C.L. and $68.340_{-1.56}^{+1.60}$ km s$^{-1}$ Mpc$^{-1}$ at 95$\%$C.L., which are almost the same with values in the $w$CDM, $w_0w_a$CDM models and improves slightly over the $\Lambda$CDM model. 
Our results are consistent with Ref.~\cite{Chen:2016uno}, which reads $H_0=68.3^{+2.7}_{-2.6}$ km s$^{-1}$ Mpc$^{-1}$ at 68$\%$ C.L. in the $\Lambda$CDM model, and also in accord with some other $H_0$ estimates in Ref.~\cite{Rigault:2014kaa,Blum:2020mgu,Freedman:2020dne,Birrer:2020tax}.
They favors the Hubble constant from Planck collaboration which is $H_0=67.4 \pm 0.5$ km s$^{-1}$ Mpc$^{-1}$ at 68$\%$C.L~\cite{Aghanim:2018eyx}, and do not prefer $H_0=74.03 \pm 1.42$ km s$^{-1}$ Mpc$^{-1}$ from Hubble Space Telescope (HST)~\cite{Riess:2019cxk}, as well as the results of Ref.~\cite{Zhang:2017aqn,Dhawan:2017ywl,Fernandez-Arenas:2017isq}.
These indicates that the $w_n$CDM models are almost impossible to solve the Hubble tension between Planck and HST thoughly. 
Moreover, we show the comoving Hubble parameter $H(z)/(1+z)$ as a function of redshift $z$ in Fig.~\ref{fig:Hz}.
The evolution of $H(z)/(1+z)$ in the $w_2$CDM model is almost the same with that of the $w$CDM model. This is resonable because $w_1$ and $w_2$ have a large overlap range in the $w_2$CDM model as shown in Fig.~\ref{fig:1d}.
And it has a slight difference from the $w_0w_a$CDM model. The difference is too small to be observed.
Thus we can not differentiate these models via the observable effects of $H(z)$.
Fig.~\ref{fig:Hz} shows the onset of acceleration in the $w_2$CDM model is around $z_\text{T}=0.6$, which is consistent with the result from Planck 2018 TT,TE,EE$+$lowE$+$lensing in the base $\Lambda$CDM model and Ref.~\cite{Haridasu:2018gqm}.
We also show the other two transition redshifts estimated from different measurements: $z_{\text{T}}=0.64^{+0.12}_{-0.09}$ from combined data of SNIa, BAO and Cosmic Chronometers (CC) data at low redshifts~\cite{Haridasu:2018gqm}; $z_{\text{T}}=0.72\pm 0.05$ from 38 measurements of $H(z)$ between redshifts $0.07 \leq z \leq 2.36$~\cite{Farooq:2016zwm}.

\begin{figure}[]
\begin{center}
\includegraphics[scale=0.3]{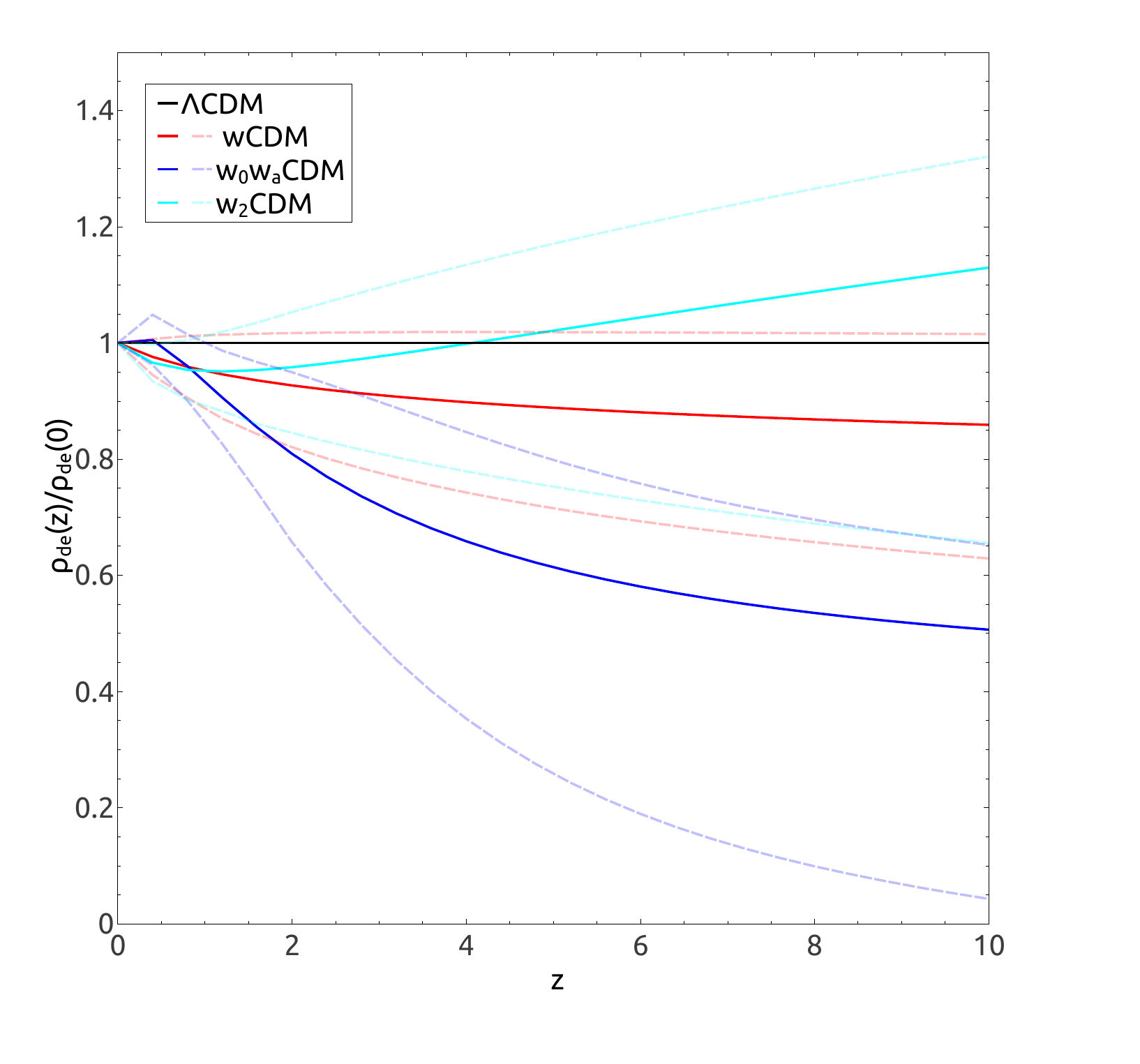}
\end{center}
\caption{$\rho_{de}(z)/\rho_{de}(0)$ as a function of redshift $z$. The horizontal black line with value of 1 indicates the values in the base $\Lambda$CDM model. The red and blue lines represent the values in the $w$CDM model and $w_0 w_a$CDM model. The cyan lines denote those in the $w_2$CDM model. The dashed lines represent their upper and lower limits of 68$\%$ C.L..}
\label{fig:rho_de}
\end{figure}

As shown in Fig.~\ref{fig:rho_de}, we normalize $\rho_{de}(z)$ in the $\Lambda$CDM, $w$CDM, $w_0w_a$CDM and $w_2$CDM models with their own $\rho_{de}(0)$.
The parameterizations of $w$CDM model and $w_0 w_a$CDM model provide almost monotonous DE density evolutions with redshift obviously.
Contrastly, the lines cross over the standard line from bottom to top in the $w_2$CDM model. They have a tick-like density evolution. This means DE in the $w_2$CDM model can make a contribution to both of late and early universe.
Thus it is possible to differentiate the $w_2$CDM model from the $w$CDM model with other measurements.

\section{Summary}
\label{sec:sum}
In this paper, we try to investigate the multicomponent DE cosmological observations.
New models named $w_n$CDM ($n=2,3,5$) models are constructed assuming DE is composed of several equal parts with individual constant EOS $w_i$.
The background and perturbation evolutions of DE are modified in the CAMB$+$CosmoMC packages.
We also modify the ``halofit" code included in the CAMB package because DE perturbations can cluster and influence the structure formation.
Then we put constraints on parameters in the $w_2$CDM, $w_3$CDM and $w_{5}$CDM model from Planck 2018 TT,TE,EE$+$lowE$+$lensing, BAO data and PANTHEON samples.
According to our results, the $w_2$CDM model is more favoured over the $w_0w_a$CDM model and $\Delta \chi^2_{\text{min}} = \Delta \text{AIC} = \Delta \text{BIC} =-2.48$.
The $w_3$CDM and $w_5$CDM models are disfavored due to their larger values of AIC or BIC.
However, the allowed ranges of the highest values of $w_i$s overlap with the lowest ones in the $w_2$CDM and $w_3$CDM models at about 2$\sigma$. When values of $n$ get larger, the maximum and minimum of $w_i$s in the $w_n$CDM models do not overlap, but the $\chi^2_{\text{min}}$, AIC, BIC also increase.
In summary, we find no evidence of multicomponent DE in the $w_n$CDM ($n=2,3,5$) models.
Moreover, the $w_2$CDM and $w_3$CDM models fit observations better than the $w_0w_a$CDM model when $\chi^2_{\text{min}}$ and AIC are considered. We find an inspiration that the $w_n$CDM models may be better than the dynamical DE models with the same numbers of free parameters.

In addition, our results show that the transition redshift in the $w_2$CDM model is around $z \sim 0.6$ and the $w_n$CDM models can relieve the Hubble tension slightly, but cannot solve it thoroughly.
Besides, the plot of total DE energy density evolution indicates that the $w_2$CDM model has a tick-like density evolution which can make a contribution to both late and early universe.
So we can expect that experiments related to $\rho_{de}$ will differentiate the $w_2$CDM and $w$CDM models.


\begin{acknowledgments}
We acknowledge the use of HPC Cluster of Tianhe II in National Supercomputing Center in Guangzhou. Ke Wang is supported by grants from NSFC (grant No. 12005084). 
\end{acknowledgments}
 
 


 \end{document}